**Die to wafer direct bonding of (100) single-crystal diamond thin films for quantum optoelectronics.**

Dominic Lepage[1,2] | Amin Yaghoobi[3] | Heidi Tremblay[1,2] | Dominique Drouin[1,2]

[1]*Institut Interdisciplinaire d'Innovation Technologique (3IT), Université de Sherbrooke, Sherbrooke, QC Canada, J1K 0A5;*

[2]*Laboratoire Nanotechnologies Nanosystèmes (LN2) ─ CNRS 3463, Université de Sherbrooke, Sherbrooke, QC Canada, J1K 0A5;*

[3]*Department of Physics, Concordia University Montréal, QC Canada, H4B 1R6*

**Correspondance:** Dominic.Lepage@usherbrooke.ca | Dominique.Drouin@usherbrooke.ca

## Abstract;

This work unlocks the manufacturing of nanophotonic quantum systems that exploit the unique material properties of single-crystal diamond (SCD). We achieve this by introducing a semiconductor-compatible process for the direct bonding of multiple high-quality, ultrathin diamond films onto a carrier wafer, enabling the subsequent parallel nanofabrication of optoelectronic integrated circuits. Central to this approach is a 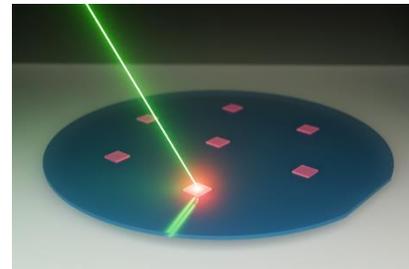 new diamond surface-preparation method that avoids boiling tri-acid mixtures while producing exceptionally clean 20 µm thin single crystals. These platelets are bonded side-by-side to 100 mm silica wafers and exhibit a record shear strength of 45.1 MPa for (100)-oriented diamond, surpassing all previously reported bonding attempts. Evidence indicates that the bonding is dominated by van der Waals interactions, likely arising from mismatched protonation mechanisms between Si–OH and C–OH surface terminations, rather than from covalent-bond-driven mechanisms. Despite this non-molecular nature, the heterostructures remain stable through liquid immersions and standard nanofabrication steps. Because the method depends primarily on surface cleanliness and roughness rather than specific chemistries, it is broadly transferable across wafer materials. This capability to parallel-bond ultrathin SCD films onto large-area substrates provides a scalable route to high-performance platforms spanning nanophotonic quantum technologies, high-power electronics, MEMS, and biotechnology.

**Introduction**

Diamonds stand out for their exceptional properties such as a wide bandgap, ultrahigh breakdown field, extreme thermal conductivity, high carrier mobility, chemical inertness, biocompatibility and remarkable mechanical strengths [1–3]. These properties would enable diamond to excel in demanding applications from high-power electronics [4] to radiation detectors [5] or MEMS devices [6]. Diamonds shine particularly bright in the current quantum technology landscape because of its ability to host stable defects like nitrogen or silicon vacancy centers. Radio-frequency manipulation and visible-wavelength optical addressing of these artificial atoms provide room-temperature readout and control of quantum states, enabling applications in quantum magnetometry, thermometry, communication, and computing [7–9]. Leveraging semiconductor nanofabrication methods to mass-produce diamond-based devices would therefore unlock major advances across multiple fields.

Because of these attributes, diamond's chemical inertness, optical transparency, and high dielectric constant can also present significant challenges for surface preparation and nanofabrication. Indeed, proof-of-concept devices on bulk diamonds substrates typically fail to maintain device-grade surface quality over footprints exceeding a few $\mu m^2$. SCD are also expensive and high-purity CVD-grown substrates are currently limited to areas of only a few mm². Polycrystalline diamond films cannot replicate the electrical, optical, or quantum properties of SCD, creating a strong need for heterogeneous bonding of SCD thin films onto larger wafers for scalable circuit fabrication. To fully harness diamond's exceptional capabilities, direct bonding to wafer substrates [10–14] is preferred over solder-based approaches [15] or fusion methods [16], because it potentially leads to significantly better mechanical and optoelectronic properties without constraining device applications.

When it comes to the development of a material bonding process, what matters at the end is the stress the heterostructure can withstand under nanofabrication and operational conditions. Various efforts have been made to establish practical strategies given the extraordinary potential offered by diamonds. Matsumae et al. [11–14,19] have done considerable work on the topic of direct diamond bonding. They found (111) diamonds to be significantly easier to bond than (100), with measured shear stress of 30-35MPa compared to 0-2Mpa respectively [14]. They attribute this major difference to the surface functional groups of (100) being mostly carbonyl groups (C=O) and ether groups (C–O–C) preventing the dehydration process capable of forming the Si-O-C covalent bonds. Unfortunately, the (111) plane presents many manufacturing challenges and most commercial CVD diamonds are grown in the (100) direction. More recently, Miyatake et al. demonstrated a (100) bonding of 14MPa on sapphire and concluded that *"[…] the surface must have a roughness of <0.2 nm and unevenness of <300 nm"* for a successful bonding [10]. When it comes to thin diamond films, the work of Yushin et al [16] illustrates how μm-scale *polycrystalline* diamond layers can be successfully bonded using high temperature fusion (950ºC). While easier to produce, polycrystalline diamonds are limited in their properties compared to single crystals, especially in fields of quantum optoelectronics. Chretien et al. [15] attempted crystalline Smart Cut transfer but had to resort to metallic soldering and ended up with significant layer graphitization, again limiting potential applications. Table 1 presents a summary of the existing body of work and their mechanical properties.

Table 1: Bonding of diamond on a functional substrate

| Authors | Diamond type | Size [mm] | Method | Shear Stress [MPa] |
|---|---|---|---|---|
| Matsumae [14] | 111 | 3x3 x0.3 | Direct | 35 |
| Matsumae [14] | 100 | 5x5 x0.5 | Direct | 0 |
| Matsumae [13] | 100 | 5x5 x0.5 | Direct | 2 |
| Fukumoto [12] | 111 | 3x3 x0.3 | Direct | 31 |
| Matsumae [11] | 111 | 3x3 x0.3 | Direct | 9 |
| Miyatake [10] | 100 | 4x4 x0.5 | Direct | 14 |
| Yushin [16] | poly | 3x4 x0.025 | Fusion | 32 |
| Chretien [15] | 100 | 4x4 x0.5 | On solder | N/A |
| **This work** | **100** | **1x1 x0.02** | **Direct** | **45** |

In this work, we demonstrate the simultaneous direct bonding of multiple SCD (100) thin films onto a 100mm double-side-polished fused silica wafers. Fused silica is selected for its compatibility in large-scale photonic quantum systems: low refractive index, ultrahigh purity free of luminescent or magnetic contaminants and transparent on all sides for ease of photonic addressing. We present a fabrication-friendly batch surface-preparation method, followed by a mechanical characterization showing shear strengths exceeding 45 MPa. Finally, we provide clear evidence of the underlying bonding mechanism and show that this approach is fully transferable to a broad range of carrier wafers.

## Results

### Diamond platelets batch preparation

In the presented work, we start with a 3x3mm x500µm quantum grade CVD grown SCD which are diced and polished into 1x1mm x20µm thin plates (i.e. platelets). From a single CVD-grown crystal, we obtain 27 usable 1mm$^2$ areas, each capable of hosting multiple photonic devices depending on intended applications. This approach maximizes wafer coverage while minimizing the use of costly single crystal diamonds. It also reduces processing efforts required in subsequent thinning steps to reach the sub-micron thicknesses required in some optoelectronic applications. In this work we use nitrogen doped diamonds because the immediate intended devices are for commercial quantum magnetometry purposes. These diamonds are well-suited due to their high photoluminescence intensity and long spin coherence times.

The process of laser dicing leaves carbonated films on the surfaces, glue and particles. The unwanted graphitic and pyrolytic domains of carbon are particularly difficult to remove using most chemical treatments. The current standard to remove these residues are Brown's 210ºC boiling nitric, perchloric and sulfuric mixture (the tri-acid method). This procedure is incompatible with existing semiconductor infrastructures due to *significant risk to the researcher and facility due to the corrosiveness, acid fuming from heating, and fire potential* [17]. Retrieving small transparent diamond platelets from such mixture would also be a very difficult manoeuvre. In addition, from our experience, this kind of acid treatment can leave contamination salt traces on the diamond surfaces.

Instead, we have developed a batch cleaning process that can be done in a low-risk environment with minimal safety protection equipment while delivering superior results. First, we sonicate in acetone to remove all the potential adhesive residues. Then, the SCD platelets are immersed in a commercial polishing compound composed of finely dispersed aluminum oxide (nominal abrasive size: 50nm pH = 8.5), which systematically removes the graphitic layers. To eliminate any remaining particles, we use a 1% solution of commercial anionic detergent for ultrasonic cleaning (pH = 8.5) and rinse thoroughly with DI water. The detergent surfactant-based wetting and emulsification agents not only removes residual particles but also reduce the surface tension, rendering the SCD slightly hydrophilic. Furthermore, the chelating/sequestering agents ensure there are no iron or other ions left which would be problematic for quantum applications. We process these platelets in parallel batch to significantly reduce the preparation time. As illustrated in Figure 1, this method produces near impeccable surfaces at large scale, essential for high-quality direct bonding and reproducible nanofabrication.

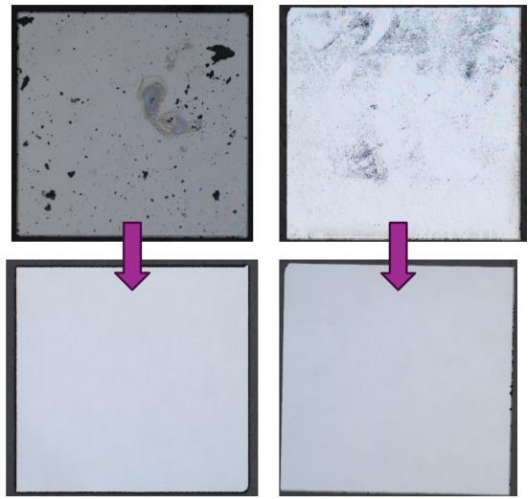

*Figure 1: Cleaning of single crystal diamond platelets. Major sources of contaminants are adhesive leftovers, surface carbonation and adsorbed particles from the environment. Our acid-free process removes everything on the 1x1mm x20μm films prior to bonding.*

**Bonding Process**

We use a commercial aligner-bonder system for reproducible, controlled and the simultaneous die bonding of multiple SCD on a single wafer. Previous work [10–14] has put forth the role of functionalization of diamonds and $SiO_2$ surfaces to promote the condensation polymerization reaction leading to covalent molecular bonding: Si-OH + OH-C ⇋ Si-O-C + $H_2O$. We will demonstrate in the subsequent sections as to why this might be a misinterpretation. Nonetheless, we prepared three types of surface functionalization: A) Deactivated, B) Indirect oxygen plasma and C) UV ozone. These methods were selected over conventional sulfuric acid activation, which proved unsuitable for producing residue-free surfaces on small SCD platelets. The "Deactivated" method consists in the heat-treatment of silica to ensure no silanol groups (Si-OH) are left, which should completely prevent any potential dehydration process. The "Indirect oxygen plasma" is done in-situ of the bonder system, where oxygen ions diffuse from a surrounding plasma ring and the bonding surfaces are not in direct contact with the energetic plasma (see methods). The UV ozone method involves exposing the surfaces to a commercial mercury vapor grid UV lamp system operated in air. Details on surface functionalization conditions are given in the methodology section.

To confirm the effects of surface activations, we relied on contact angle measurements performed on separate diamonds and silica wafers. First, on the deactivation of silica surfaces, the dominance of siloxane groups (Si-O-Si) was confirmed with average contact angles of 73±1°. Once activated, the $SiO_2$ became hydrophilic with contact angles of 4±1° for the indirect plasma and 0±1° for the UV ozone as expected for good silanol surface coverage. We measured the CVD-grown diamonds to be hydrogen-terminated with 58±5° contact angles. The commercial anionic detergent left the surfaces slightly hydrophilic with 23±2° and following surface treatments angles went down to 16±4° and 6±1° for indirect plasma and UV ozone respectively. These values are summarized in Table1 and are coherent with the work of Zulkharnay [21] and Li [22] to confirm the strong coverage of hydroxyl terminations on the SCD prior to bonding. Note that the indirect plasma approach does seem to be a valid method of C-O-C bond hydrolysis, despite the lower kinetic energies involved compared to the direct plasmas employed by Zulkharnay and Li.

**Table 2:** Summary of surface activations using contact angle measurements.

| Surfaces | Contact Angles [deg] |
|---|---|
| **SCD - Initial** | 58 ± 5 |
| **Fused Silica - Initial** | 21 ± 1 |
| **SCD -Cleaned** | 23 ± 2 |
| **Fused Silica - Deactivated** | 73 ± 1 |
| **SCD -UVO3** | 6 ± 1 |
| **Fused Silica - UVO3** | 0 ± 1 |
| **SCD -Plasma** | 16 ± 4 |
| **Fused Silica - Plasma** | 4 ± 1 |

Details of the bonding process are presented in the methodology section, but the general procedure is as follows: the diamond platelets are positioned on a bottom wafer platen and the receiver wafer on a top platen. Vacuum is made in the chamber and plasma surface activation is done in-situ when applicable. Water vapours are injected at low pressure to draw the two surfaces together, compensate for warps and initiate polymerization [23,24]. While AFM measurements of both the diamond platelets and the silica wafer present acceptable surface roughness of RMS = 0.61 ±0.08nm and 0.37±0.10nm for diamond and $SiO_2$ respectively. Water bridging enable the accommodation of larger total surface roughness 10 Å for the potential polymerization process

[24]. The platens are then joined together to apply a small uniform pressure of 50kPa. The bonded system is heated in-situ to evacuate the $H_2O$ molecules. When bonding single crystal diamonds, or multiple platelets simultaneously, the parallelism and thickness variations of the crystals can lead to cleaving of the receiver wafers. To prevent this issue, the diamonds are placed on a PTFE gasket that limits the pressure between the diamonds and the top wafer. Finally, the wafer is placed in a desiccator for 72h and a slow 24h 250ºC thermal annealing cycle is done to promote bond conversion. Figure 2 presents an example of three diamonds thin films bonded in parallel on a silica wafer.

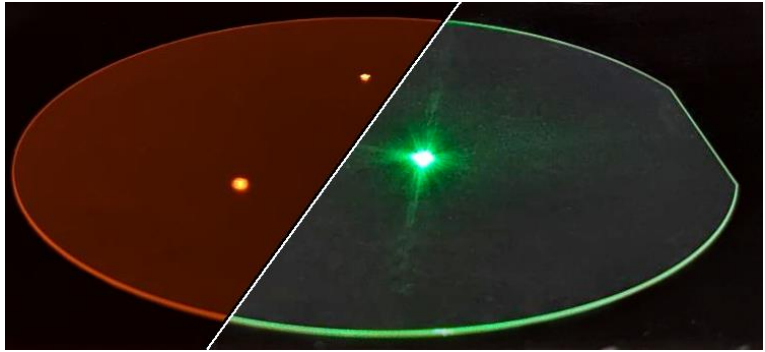

*Figure 2: Bonded quantum-grade diamond films. Resulting fused silica wafer onto which three 20µm-thick single crystal diamonds films have been bonded as an example. A laser pointer illuminates the platelets: On the right side is a normal picture and on the left side is the same wafer viewed through a red filter.*

**Shear stress measurements**

Figure 3 presents the shear stress of the diamonds that were bonded using the different surface activation techniques. Several conclusions can be drawn from this figure: First, we measured shear strengths up to 45.1±0.6 MPa, which is by far the highest adhesion strength recorded for any diamond's single crystals. This is observed for the in-situ indirect plasma activation methods. The UV ozone method presented a lower shear strength of 34±0.4 MPa and the Deactivated interfaces still presented 30±0.4MPa, which is still significantly above the best work done on (100) diamonds thus far. The fact that the deactivated surfaces yield such strengths brings doubt to the surface dehydration hypothesis for diamond bonding. We also observe no break in the force diagram of Figure 3 (and visually under microscope); the diamond is *dragged* on the surface at 30-45MPa and not broken as would be expected by a molecular bonding. As in Fukumoto et al [12], we did observe one diamond cleaved by the procedure, but each piece still presented half the shear stress of the initial one.

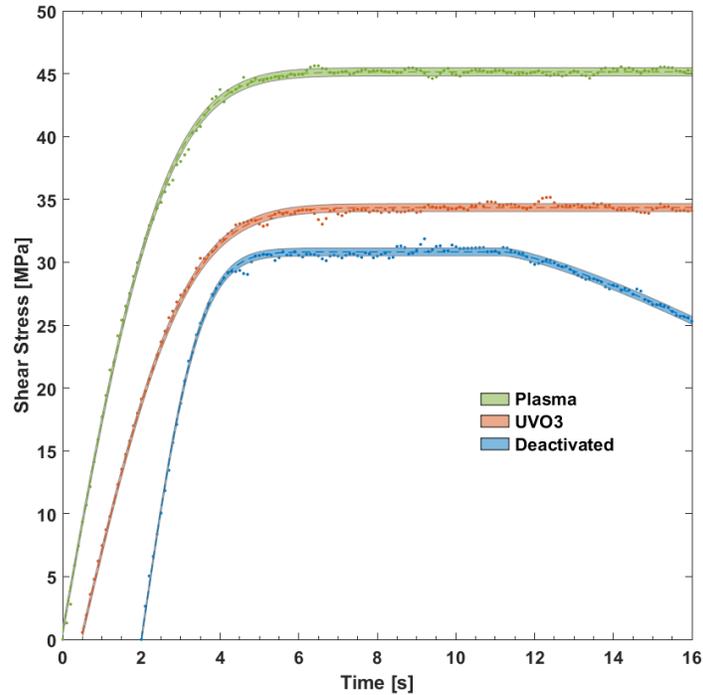

*Figure 3: Shear stress vs time curves. Three surface activation method were tested: Plasma, UV ozone and Deactivated. The constant force plateau is a first indicator that bonding is not molecular in nature. The widths of the colored curves represented the error on the curve fitting.*

Finally, we placed a drop of pure isopropyl alcohol on the SCD and measured a significant decrease in the shear stress measurements. In addition to alcohol, we also tested adhesion under various liquids commonly used in our nanofabrication processes by placing the bonded SCD in various liquids for a 15-minutes sonication: Oxylene, Methyl isobutyl ketone, ZED-N50, ZED-N50 developer and 1% detergent in $H_2O$. In all these cases, the SCD did not move, but the adhesions strength was reduced to a point where the SCD could translate on the wafer surface under strong mechanical stress such as a high-powered water jet. All these elements indicate a bonding that is van der Waals in nature and not molecular as was hypothesised by previous authors. A summary of the results is presented in Table 2.

Table 3: Measured shear stress for different surface activations methods

|  | Dry Shear [MPa] | IPA Shear [MPa] |
|---|---|---|
| SCD - Initial | 0 | 0 |
| Fused Silica - Initial | | |
| SCD -Cleaned | 30.8 ± 0.4 | 18.0 ± 0.4 |
| Fused Silica - Deactivated | | |
| SCD -UVO3 | 34.3 ± 0.4 | 17.7 ± 0.4 |
| Fused Silica - UVO3 | | |
| SCD -Plasma | 45.1 ± 0.6i | 12.7 ± 0.4 |
| Fused Silica - Plasma | | |

**Adhesion forces**

To explain the variations in shear strengths within surface treatments and the literature, we turn to Lifshitz's theory of molecular attractive forces between solids [25] to estimate the van der Waals forces between diamond and silica surfaces, both normal and transversal. The calculation of these forces, which are detailed in the SI, are inversely proportional to the gap between the two surfaces squared ($d^{-2}$) and proportional to the nonretarded Hamaker constant ($A_H$) of two surfaces interacting across a third medium [26]. In air (or vacuum), $A_H \approx 130 \times 10^{-21}$ J which decreases to $28 \times 10^{-21}$ J in $H_2O$ and $16 \times 10^{-21}$ J in IPA. Israelachvili estimates shear stress by calculating the change in surface energy per unit of lateral displacement as the top surface geometry is being dragged across the bottom one [26]. We execute the same procedure by calculating the expected value of the work at rest compared to when the SCD is moving. Figure 4 presents the theoretical shear stress as a function of surface roughness between diamond and silica. The shaded green region represents the observed experimental shear stress, and it falls surprisingly close to the measured values as a first approximation. Further details on these calculations are provided in the SI. We believe the $d^{-2}$ relationship, which is a direct function of the quality of the diamond cleaning and polishing, is enough to explain the variations observed in Table 3 and Figure 3.

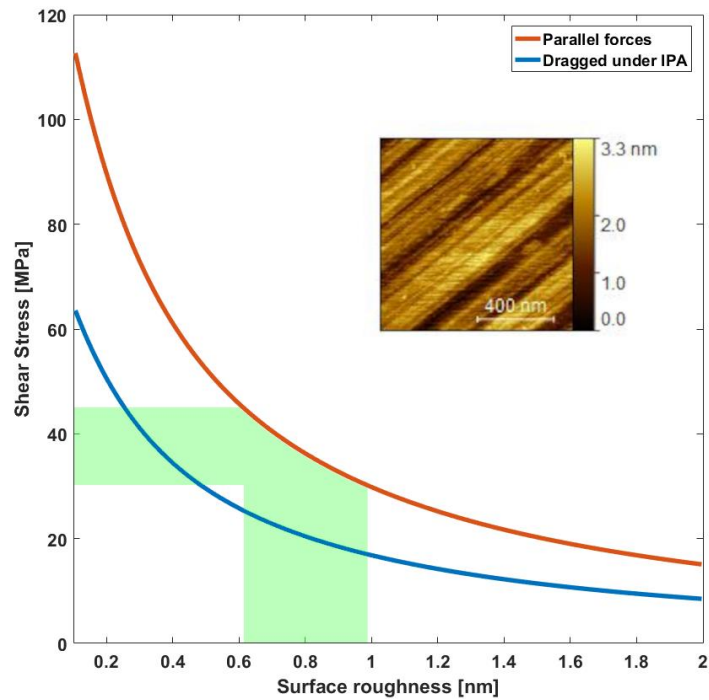

*Figure 4: Theoretical shear stress calculation. Using the Lifshitz's theory of molecular attractive forces we can estimate the shear stress of two surfaces bounded only by Van der Waals forces. In shaded green, the roughness zone we operate under leading to 30-45MPa. Insert: AFM of the polished diamond platelet used in the calculations.*

Finally let us examine the forces involved in a potential Si-O-C molecular bonding: The surface energy of the Si-O-C bond, is estimated to be at least 2.74 $J/m^2$ [27–29] for experimental silanol coverage and annealing temperatures [23,24]. Breaking such bonds would lead shear stresses measurements of >17GPa. This value is orders of magnitude higher than any experimental data observed in Table 1, which bring further doubt to the molecular bonding hypothesis. More details and experimental references on this estimate are provided in the SI.

## Discussion

All indicators lead us to believe that this bonding is Van der Waals in nature and not molecular: i) The shear stress profile versus time does not present a break: the SCD are *dragged* on the surfaces at 30-45.1 MPa ii) Deactivated and activated surfaces have comparable shear stress iii) Immersing the SCD in liquids reduce their shear stress iv) Estimating the shear stress theoretically using measured AFM roughness leads to comparable forces and finally v) The theoretical strength of molecular bonding should be several orders of magnitude higher than our record shear stress measurements.

There are three potential reasons for which we do not seem to observe molecular bonding between diamond and silicon/silica. First, could the surface roughness be too large to allow for such bonding? Supposing the distance between the two interface (d) is randomly distributed, the average distance between the two interfaces is $\bar{d} = 4.8$Å and can vary on the surface between 0.96Å (confined water [30]) to > 12.9Å (polishing grooves are maximally opposed). The formation of a Si-O-C covalent bond should necessitate the surfaces to be within 2.5Å [31], but could be accommodated up to 10Å according to DFT theoretical models and experiments [24,30,31]. We conclude that a significant fraction of the diamond – silica interface should be within proximity for direct bonding. Second, could it be that $H_2O$ is not removed adequately during the desiccation and annealing? In the case of homogeneous silicon/silica bonding, experiments show that this procedure is more than adequate to diffuse $H_2O$ across large-diameter wafers [23] and for the healing of cracks in bulk silicate [29]. It is therefore unlikely that mm-scale diamonds of a larger RMS than standard wafers would leads to water pockets at the interface.

A more plausible explanation is that C–OH groups simply cannot substitute an Si–OH groups in hydroxyl-mediated polymerization. To understand why, consider the better known and simpler case of molecular bonding in silicon and silica surfaces. Two types of silanol groups exist on Si/$SiO_2$ surfaces: out-of-plane silanols with strong acidic character (pKa = 5.6) and more alkaline in-plane silanols (pKa = 8.5). The former form short, strong hydrogen bonds with interfacial water molecules, whereas the latter establish weaker hydrogen bonds [24,28,30]. For molecular bonding to occur, associated silanol groups must dominate over isolated ones. Upon gentle heating, increased surface mobility promotes the formation of the more energetically favorable, hydrogen-bonded interfacial water structure. Before annealing, the partially ordered silanol configuration yields a surface energy of approximately 0.33 J·m$^{-2}$. Under these conditions, protons transfer to the neighbouring OH group is promoted because $pKa_{associated} = 5.6 < pH = 7 < pKa_{isolated} = 8.5$ and $H_2O$-based polymerization can take place. For diamond surfaces, however, two mismatches prevent analogous behavior: First, Zilibotti et al. reported that diamond surface hydroxyls are predominantly in-plane [32], which structurally isolates C–OH groups from the bonding configuration observed in the Si/$SiO_2$ system. Second, Chakrapani et al. measured the diamond hydroxyl pKa to be approximately pKa ≈ 10 [33], indicating a highly alkaline character. As shown in Figure 5, while silanol groups become increasingly prone to molecular bonding during mild heating and deprotonation, the diamond hydroxyls remain highly protonated and therefore cannot polymerize with adjacent $H_2O$ chains, even when sterically allowed. Under a more alkaline environment, perhaps the isolated C-OH could move and deprotonate, but then the silicon/silica surface would be composed of a mixture of isolated and associated silanols.

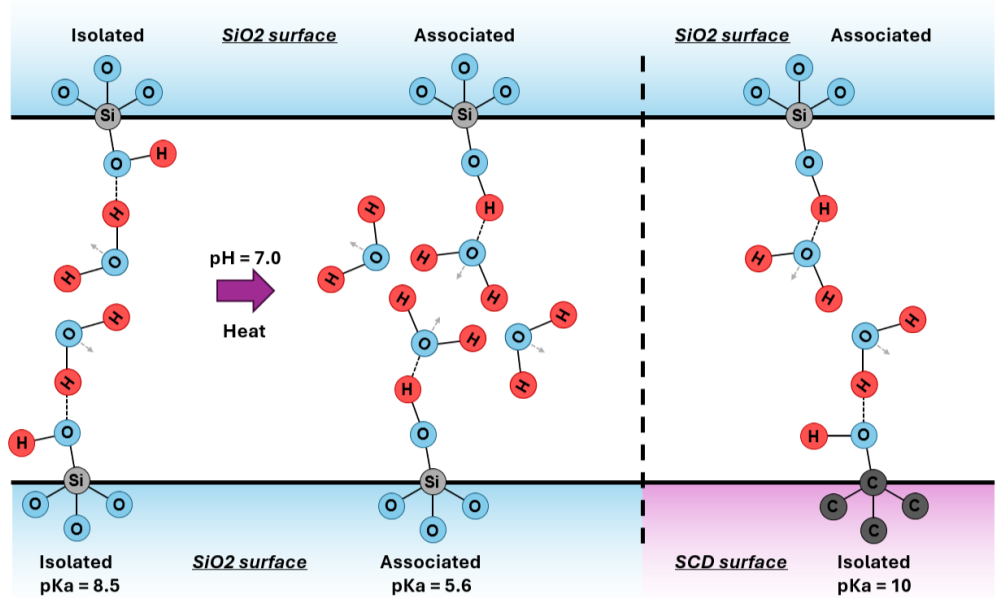

*Figure 5: Water bridging hydrophilic silica and diamond surfaces.* *For the molecular bonding of silicon or silica, a small amount of heat suffices for water to gain increased surface mobility and form energetically favorable structures for molecular bonding. For the diamond surface, this cannot be the case due to the orientation of the hydroxyl and the pkA of the C-OH termination.*

In conclusion, we have successfully demonstrated a novel method of single crystal diamond (SCD) surface preparation which does not involve the use of boiling tri-acid mixtures. We use this method to prepare very clean and thin 1x1mm x20μm diamond platelets which are then bonded in parallel on the surfaces of 100mm silica wafers. Shear-stress measurements of these (100) diamond films present a record strength of 45.1 MPa which is significantly beyond any other previous attempts. The bonding seems to be Van der Waals in nature and not resulting from molecular interactions between the two surfaces. A plausible explanation for this is the mismatch in protonation mechanisms for Si-OH and C-OH. The resulting bond is however strong enough to survive liquid immersions and the subsequent nanofabrication processes. Because the process is more dependent on surface cleanliness and roughness than its chemistry, we have also successfully used this method with other flat substrates such as silicon and sapphire wafers. Given the outstanding properties of SCD, we believe the parallel bonding of such thin films on wafers will enable the mass production of systems of exceptional and novel performances in the fields of nanophotonic quantum, high-power electronics, MEMS and biotechnology.

## Methodology

### Cleaning process:

In the presented work, the diamonds start as 3x3mm x500μm from Element Six's general-purpose quantum grade CVD single crystal diamonds (SCD). The diamonds were then diced and polished into 1x1mm x20µm platelets by Applied Diamond Inc. To manipulate the SCD platelets between steps, we use a vacuum pick-up pen with a 45° stainless steel blunt needles and 650µm opening.

After an initial optical inspection using a Keyence confocal microscope, the platelets are placed individually in porcelain Gooch crucible filters with 800µm holes at the bottom. This offers chemical resistance and a capacity to substantially rinse between steps. The platelets are placed in separate filters to avoid them sticking to each other. An initial 30min sonication at 37kHz in warm acetone is done using an Elmasonic P system to remove glue residuals. To remove the carbonated films on the diamond's surfaces, they are submerged in a MasterPrep polishing suspension of 50nm alumina and sonicated at room temperature for 50minutes. The polishing is completed by an exhaustive rinsing process using DI water and more sonication to dislodge most adsorbed particles. The SCD are then transferred to a new set of Gooch filters dedicated to cleaning and are submerged in a 1% Liquinox anionic detergent at 70°C. We sonicate again for 25 minutes, this time agitating the filters to prevent the SCDs from sticking to one face of the filter. A thorough rinsing process using 70°C DI water is done in the same Elmasonic sonication system. Our 2.8L tank could accommodate 8 filters at a time. At the end of the rinsing process, the SCD are transferred to a PTFE plate, a quality control is made at the confocal microscope and the PTFE is moved to the bonding station.

**Activation and bonding processes:**

The diamond and silica surfaces were inspected using a Bruker's VERTEX 70v FTIR and a Seagull variable angle reflection accessory by Specac Inc. The various methods of ATR measurements are detailed in the SI. Contact angle measurements were done using a DSA30S drop shape analyser by Kruss.

Regarding the surface activations methods: a) For the 'Deactivated' interface, the silica wafer was heated at 350°C for 2hrs in a Carbolite Gero controlled atmosphere oven. This turns a silanol silica surface (contact angle $0 \pm 1°$) into a siloxane-dominant surface (contact angle $73 \pm 1°$). Nothing was done to the diamond platelets other than the cleaning sequence described above. b) For the UV ozone, the silica and diamonds were placed separately in a Novascan Ozone PSD-UVT surface system for 10minutes at 65°C. c) The indirect plasma is generated by the in-situ radical activation system of an AWB04 wafer bonding machine from Applied Microengineering Ltd (AML). This equipment exposes diamonds and wafers simultaneously to elemental O along with $O^+$ and $O_2^+$ radicals. We run this indirect plasma for 60minutes. The 1-hour benchmark was established as the time required to uniformly turn a siloxane-dominant wafer (contact angle $73 \pm 1°$) to a silanol one (contact angle $4 \pm 1°$) across the 100mm diameter. As mentioned in the discussion section, we believe all these methods are equivalent for Van der Waals bonding, which is more dominated by surface roughness than surface activation.

Nonetheless, the SCD platelets on PTFE are placed on the bottom platen of the AWB04 system while the top platen holds the substrate wafer. Indirect plasma activation is done if applicable. Following this, water vapour is injected in-situ at 10 mbar using AML valve system and left to react with the surfaces for 5 min. The parallel platens are then closed under a small uniform pressure of 4 MPa (shear strength of PTFE). We anneal under pressure at 100°C for one hour following a slow temperature ramp of 1°C/min from 21°C to 100 °C. This process minimizes void formation in silicon wafer bonding processes and promotes the formation of short hydrogen bonds chains between silanols. The wafers were then taken out of the AML system and placed in a desiccator for 72hours to ensure that no moisture was left. After this waiting period, we followed the established procedure for silanol polymerization by annealing in a Carbolite Gero controlled atmosphere oven. The temperature ramps were 100°C at 1°C/min, then 250°C at 2°C/min, annealing at 250 °C for 24hrs and then a 24hrs cool-down to room temperature. A quality control was made at the confocal microscope.

**Shear stress measurements:**

The shear stress measurements are made using a Nordson Dage 4000Plus system using a 50kg load cartridge. A modified silicon carbide coated blade was used to avoid slicing the standard steel blade by the thin diamond edges. The uncertainties presented in the Result section come from the error on the curve fitting of the data under constant dragging forces (plateaus in Figure 3).

During the process, one of the indirect plasma-activated diamond cleaved in half, but the two parts stayed with a few hundred microns of each other. We were able to do a shear stress on the pair and the resulting measurement went from 45.1 MPa to 43.1 MPa.

## Acknowledgements

The work at the Université de Sherbrooke was financed by the Quantum Sensor Challenge (QSP111) in partnership between the National Research Council and SBQuantum Inc. In addition, this work was funded by the Mitacs Innovation QSciTech Program and the Laurent and Claire B. Beaudoin scholarship.

## Author contributions

D.D. and D.L. conceived the idea. A.Y. developed the initial AML wafer bonding procedure with D.L. H.T. and D.L. developed the cleaning process, did the diamond bonding and the surface characterization measurements. D.L. conducted the shear stress measurements, interpreted the data and wrote the manuscript. D.D. supervised the research.

# Supplementary information
Die to wafer direct bonding of (100) single-crystal diamond thin films for quantum optoelectronics.

Content:

**Note1:** Calculation of shear stress

## Note 1: Calculation of shear stress

**Van der Waals interfaces**

When dragging two surfaces against one another, the friction is determined by the load and adhesion between them. There is no compressive load in our measurements ($F_\perp = 0$) and we measure the behaviour as a function of $F_{ll}$ only. Van der Waals adhesion-controlled friction is driven by the surface geometry: At rest, the average surface energy $W_0$ is maximised by the bonding process. As illustrated in Figure S1, when dragging the top surface (SCD) across the substrate a certain distance $\Delta x$, the roughness must clear the bottom roughness and in doing so is changing in height $\Delta z$. This leads to a reduction of the interface energy $W_1 < W_0$. If we neglect the kinetic energy losses from roughness impacts [1], the shear stress is then given by the change in surface energy per unit distance as described in Equation S1:

$$F_{shear} = \Delta W(z) / \Delta x \quad \text{(Eqn. S1)}$$

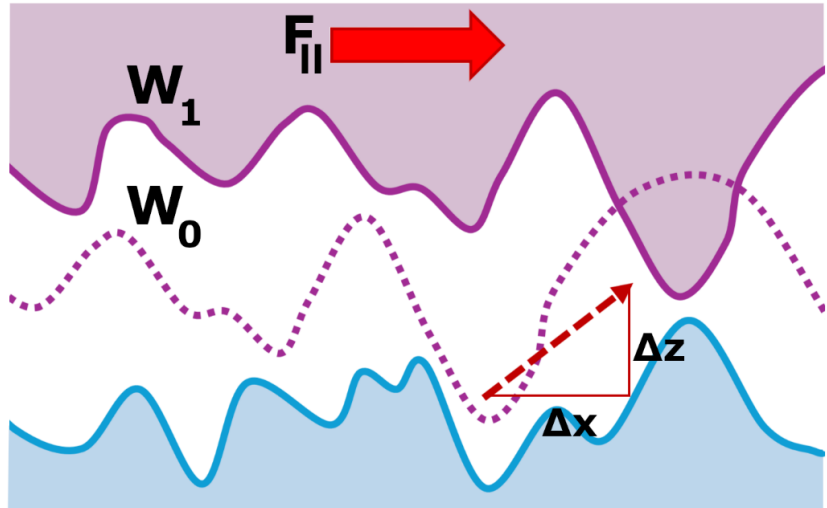

**Figure S1 Lateral force (shear) emerging from roughness.** The normal ($\Delta z$) and lateral ($\Delta x$) displacement of the top surface changes the interface energies from $W_0$ to $W_1$.

The scale of Figure S1 is greatly exaggerated for clarity: The two surfaces are very flat and using the AFM measurements we estimate their fractal dimensions to be $(2+2\times10^{-5}) \approx 2$. For this reason, we can use equations that describe two flat surfaces [2]. Van der Waals interaction energy is given by Equation S2.

$$W = -A_H / 12\pi d^2 \quad \text{(Eqn. S2)}$$

Where d is the separation between the two surfaces and $A_H$ the nonretarded Hamaker constant given by Equation S3:

$$A_H \approx \frac{3}{4} k_B T \left(\frac{\epsilon_1 - \epsilon_3}{\epsilon_1 + \epsilon_3}\right)\left(\frac{\epsilon_2 - \epsilon_3}{\epsilon_2 + \epsilon_3}\right) + \frac{3h\nu_e}{8\sqrt{2}} \frac{(n_1^2 - n_3^2)(n_2^2 - n_3^2)}{(n_1^2 + n_3^2)^{1/2}(n_2^2 + n_3^2)^{1/2}\{(n_1^2 + n_3^2)^{1/2} + (n_2^2 + n_3^2)^{1/2}\}} \quad \text{(Eqn. S3)}$$

In air (or vacuum), $A_H \approx 130\times10^{-21}$ J which decreases to $28\times10^{-21}$ J in $H_2O$ and $16\times10^{-21}$ J in IPA. To estimate the values for the interface energies, we approximate the surface separation d by a truncated normal distribution with $d_{min} = 0.096$ nm corresponding to a single hydroxyl group (OH) and $d_{max} = \sqrt{2} \times (RMS_{SCD} + RMS_{Silica})$ with the RMS measured at the AFM. We can then take the expected first and second moments of Equation S2 where d is distributed as such to uncover the average surface energy $\mu_W = E[W] = 5.4$ mJ/m² and its standard deviation $\sigma_W = \sqrt{(E[W^2] - E[W]^2)} = 10.8$ mJ/m² for our configuration.

The correlation length of a rough (or periodic) surface is a statistical measure that quantifies the lateral distance over which surface height variations remain self-similar [3]. The displacement Δx is ½ of a correlation length, i.e. approximately the distance between peaks and troughs across the surface. This can be calculated by measuring the decay of the cross-correlation of the AFM measurements which is Δx = 0.54nm in our case.

Taking the variations in energies to be ±$\sigma_W$ as the top surface is being dragged per unit Δx, Equation S1 yields $F_{Shear}$ ~40 MPa, which is approximately what we measure. The behaviour of equation S1 as a function of the surface roughness of the SCD is illustrated in Figure 4 of the main document. We notice how a slight variation in the surface roughness or cleanliness will lead to large differences in $F_{Shear}$ values.

**Molecularly bonded interfaces**

To estimate the shear stress involved in a molecularly bonded interface, equation 1 is still used. However, instead of a Van der Waals calculation, we use tabulated values for the dissociation of bond energies. The surface density of the hydroxyl is $d_{OH1} = 1.4 \times 10^{18}$ m⁻² and $d_{OH2} = 3.2 \times 10^{18}$ m⁻² for the isolated and associated group respectively. Their hydrogen bonded energy is $E_{H1} = 0.43$ eV and $E_{H2} = 0.26$ eV respectively. The specific surface energy is thus $W_0 = (2 \times d_{OH1} \times E_{H1} + d_{OH2} \times E_{H2}) = 0.33$ J/m² which is slightly below what has been measured for unannealed siloxane surfaces [4]. Following the annealing and molecular bonding, the surface density of the bonds is $d_{OH1} + d_{OH2} = 4.6 \times 10^{18}$ m⁻². The energy required to break a siloxane bond (Si-O) is 4.7eV and 3.7eV for carbon monoxide (C-O) [5]. Therefore, the molecularly bonded surface energy is $W_{Si-O} = 3.45$ J/m² and $W_{C-O} = 2.74$ J/m². If we were to break these bonds over distances of one molecular length, that is Δx = 0.163nm for Si-O and Δx = 0.143nm for the C-O bonds, the measured shear strengths would be $F_{Shear\,Si-O} = 19$GPa and $F_{Shear\,C-O} = 17$GPa respectively.